
%
\magnification=\magstep1
\input amstex
\vsize=22.5truecm
\hsize=16truecm
\hoffset=2.4truecm
\voffset=1.5truecm
\parskip=.2truecm
\font\ut=cmbx10
\font\ti=cmbx10 scaled\magstep1
\font\bfsl=cmbxsl10

\def\br{\hfill\break\noindent}
\TagsOnRight
\newfam\rmfam
\def\rm{\fam\rmfam}
\font\tenrm = cmr10
\font\sevrm = cmr7 \skewchar\sevrm='177
\font\sevit = cmmi7 \skewchar\sevrm='177
\font\fivrm = cmr5 \skewchar\fivrm='177
\textfont\rmfam = \tenrm
\scriptfont\rmfam = \sevrm
\scriptscriptfont\rmfam = \fivrm

\def\al{\alpha}
\def\be{\beta}
\def\ga{\gamma}
\def\de{\delta}
\def\eps{\epsilon}
\def\la{\lambda}

\def\Om{\Omega}

\def\vx{\vec{x}}
\def\vxp{\vec{x}\,'}
\def\vy{\vec{y}}

\def\vk{\vec{k}}

\def\vp{\vec{p}}

\def\vq{\vec{q}}


\def\od2{\overline{|\de_k|^2}}

\def\11r{|1,\overline{1};\Om\rangle}
\def\l11{\langle1,\overline{1};\Om|}

\def\vt{\!\triangle\!\!}

\def\lsim{\mathrel{\mathchoice {\vcenter{\offinterlineskip\halign{\hfil
$\displaystyle##$\hfil\cr<\cr\sim\cr}}}
{\vcenter{\offinterlineskip\halign{\hfil$\textstyle##$\hfil\cr<\cr\sim\cr}}}
{\vcenter{\offinterlineskip\halign{\hfil$\scriptstyle##$\hfil\cr<\cr\sim\cr}}}
{\vcenter{\offinterlineskip\halign{\hfil$\scriptscriptstyle##$\hfil\cr<\cr
\sim\cr}}}}}
\def\gsim{\mathrel{\mathchoice {\vcenter{\offinterlineskip\halign{\hfil
$\displaystyle##$\hfil\cr>\cr\sim\cr}}}
{\vcenter{\offinterlineskip\halign{\hfil$\textstyle##$\hfil\cr>\cr\sim\cr}}}
{\vcenter{\offinterlineskip\halign{\hfil$\scriptstyle##$\hfil\cr>\cr\sim\cr}}}
{\vcenter{\offinterlineskip\halign{\hfil$\scriptscriptstyle##$\hfil\cr>\cr
\sim\cr}}}}}

\def\aki{a_{k,\,{\rm in}}}
\def\uki{u_{k,\,{\rm in}}}

\def\kph{k_{\rm phys}}
\def\qph{q_{\rm phys}}

\def\laph{\la_{\rm phys}}
\def\eph{e_{\rm phys}}
\def\Hi{H_{\rm I}}
\def\ri{\rho_{\rm I}}
\def\Heq{H_{\rm EQ}}
\def\laeq{\la_{\rm EQ}}
\def\kpheq{k_{\rm phys,\,EQ}}
\def\Mp{M_{\rm Pl}}

\def\r{\langle\rho\rangle}
\def\rx{\langle\rho(x)\rangle}

\def\rt{\rho_{\rm tot}}

\def\rmk{\rho_{\raise1pt\hbox{\sevrm mode}\,\,\vk}}
\def\rmp{\rho_{\lower1pt\hbox{\sevrm mode}\,\,\vp}}
\def\drmk{\de\rho_{\raise1pt\hbox{\sevrm mode}\,\,\vk}}
\def\drmq{\de\rho_{\lower1pt\hbox{\sevrm mode}\,\,\vq}}
\def\ek{E_{\vk}}
\def\rk{\rho_{\vk}}
\def\emk{e_{\vk}}
\def\dek{\de E_{\vk}}
\def\drk{\de\rho_{\vk}}
\def\dnk{\de n_{\vk}}

\def\ddR{\langle\de_R^2\rangle}

\def\dddRS{\langle\de_R^2\de_{\lower1pt\hbox{\sevit S}}\rangle}
\def\dors{\de\overline\rho_S}

\def\ors{\overline\rho_S}
\def\or{\overline\rho}
\def\Drs{\Delta\rho_S}
\def\des{\de E_S}

\font \fivesans               = cmss10 at 5pt

\font \tenrm                  = cmr10
\font \sevensans              = cmss10 at 7pt
\font \tensans                = cmss10
\newfam\sansfam
\textfont\sansfam=\tensans\scriptfont\sansfam=\sevensans
\scriptscriptfont\sansfam=\fivesans
\def\sans{\fam\sansfam\tensans}



\def\bbbc{{\mathchoice {\setbox0=\hbox{$\displaystyle\hbox{\rm C}$}\hbox{\hbox
to0pt{\kern0.4\wd0\vrule height0.9\ht0\hss}\box0}}
{\setbox0=\hbox{$\textstyle\hbox{\rm C}$}\hbox{\hbox
to0pt{\kern0.4\wd0\vrule height0.9\ht0\hss}\box0}}
{\setbox0=\hbox{$\scriptstyle\hbox{\rm C}$}\hbox{\hbox
to0pt{\kern0.4\wd0\vrule height0.9\ht0\hss}\box0}}
{\setbox0=\hbox{$\scriptscriptstyle\hbox{\rm C}$}\hbox{\hbox
to0pt{\kern0.4\wd0\vrule height0.9\ht0\hss}\box0}}}}

\def\bbbq{{\mathchoice {\setbox0=\hbox{$\displaystyle\hbox{\rm Q}$}\hbox{\raise
0.15\ht0\hbox to0pt{\kern0.4\wd0\vrule height0.8\ht0\hss}\box0}}
{\setbox0=\hbox{$\textstyle\hbox{\rm Q}$}\hbox{\raise
0.15\ht0\hbox to0pt{\kern0.4\wd0\vrule height0.8\ht0\hss}\box0}}
{\setbox0=\hbox{$\scriptstyle\hbox{\rm Q}$}\hbox{\raise
0.15\ht0\hbox to0pt{\kern0.4\wd0\vrule height0.7\ht0\hss}\box0}}
{\setbox0=\hbox{$\scriptscriptstyle\hbox{\rm Q}$}\hbox{\raise
0.15\ht0\hbox to0pt{\kern0.4\wd0\vrule height0.7\ht0\hss}\box0}}}}

\def\bbbz{{\mathchoice {\hbox{$\sans\textstyle Z\kern-0.4em Z$}}
{\hbox{$\sans\textstyle Z\kern-0.4em Z$}}
{\hbox{$\sans\scriptstyle Z\kern-0.3em Z$}}
{\hbox{$\sans\scriptscriptstyle Z\kern-0.2em Z$}}}}
\def\qed{\ifmmode\sq\else{\unskip\nobreak\hfil
\penalty50\hskip1em\null\nobreak\hfil\sq
\parfillskip=0pt\finalhyphendemerits=0\endgraf}\fi}


\def\bla{\big\langle}
\def\bra{\big\rangle}
\def\H{\Cal{H}}
\def\x{\rm x}
\def\y{\rm y}

\def\gati{\ga^{\lower2pt\hbox{$\ssize\tilde0$}}}

\rightline{ETH-TH/94--25}
\rightline{December 1994}
\vskip2truecm
\centerline{\ti Energy Density Fluctuations in Inflationary Cosmology}
\vskip2truecm
\centerline{Harald F. M\"uller and Christoph Schmid}
\vskip0.5truecm
\centerline{Institut f\"ur Theoretische Physik, ETH--H\"onggerberg,
            CH--8093 Z\"urich, Switzerland}

\vskip3.5truecm

\centerline{\ut Abstract}
\vskip0.5truecm
\noindent{\sl
 We analyze the energy density fluctuations contributed by scalar fields
 $\Phi$ with vanishing expectation values, $\langle\Phi\rangle=0$, which are
 present in addition to the inflaton field. For simplicity we take $\Phi$ to
 be non--interacting and minimally coupled to gravity. We use normal
 ordering to define the renormalized energy density operator $\rho$, and we
 show that any normal ordering gives the same result for correlation
 functions of $\rho$. We first consider massless fields and derive the
 energy fluctuations in a single mode $\vk$, the two--point correlation
 function of the energy density, the power spectrum, and the variance of the
 smeared energy density, $\ddR$. Mass effects are investigated for energy
 fluctuations in single modes. All quantities considered are scale invariant
 at the second horizon crossing (Harrison--Zel'dovich type) for massless and
 for unstable massive fields. The magnitude of the relative fluctuations
 $\de\rho/\rt$ is of order $(\Hi/\Mp)^2$ in the massless case, where $\Hi$
 is the Hubble constant during inflation. For an unstable field of mass
 $m_\Phi\ll\Hi$ with a decay rate $\Gamma_\Phi$ the magnitude is enhanced by
 a factor $\sqrt{m_\Phi/\Gamma_\Phi}$. Finally, the prediction for the
 cosmic variance of the average energy density in a sample is given in the
 massless case.
}


\vfill\eject

\noindent{\ti 1. Introduction and Conclusions}
\vskip0.5truecm
In the early universe gravity acting on a quantum matter field [1] amplifies
energy density fluctuations above the unavoidable vacuum fluctuations [2,3].
These primordial fluctuations may serve as seeds for the subsequent
formation of the observable large--scale structure [4].

In the standard literature [5] the fluctuations in the energy density of the
inflaton field, which drives inflation through the energy density of its
vacuum expectation value, were thoroughly analyzed.

In a recent paper [6] we have started to investigate the energy density
fluctuations contributed by scalar fields $\Phi$ with vanishing expectation
values, $\langle\Phi\rangle=0$, which are present during inflation in
addition to the inflaton field. For simplicity we take $\Phi$ to be
non--interacting and minimally coupled to gravity. The emphasis of the first
paper has been on the quantum field theoretical concepts and methods. (They
are briefly reviewed in the next section.) The present paper contains the
application to cosmology. We evolve the energy density fluctuations through
the radiation era up to the time of matter and radiation equality. A related
investigation for the case of classical cosmological perturbations can be
found in ref. [7].

The energy density $\rho$ of the (non--inflaton) field $\Phi$ is bilinear in
the fluctuating field, and $\rho$ is therefore not a Gaussian variable. When
$\Phi$ is non--interacting and in external gravity, $\Phi$ is a Gaussian
variable and $\rho$ is $\chi^2$--distributed. This is in contrast to the
Gaussian energy density fluctuations for the inflaton field, which arise
from the interference between the background field and the fluctuating part
of the inflaton. Various observable measures of non--Gaussianity in our
model are discussed in ref. [8].

We use normal ordering to define the renormalized energy density operator,
and we show that any normal ordering prescription gives the same results for
correlation functions of $\rho$. This is so because the difference of two
normal orderings is a c--number, which drops out in the correlations.

The cosmological model is given as an inflationary universe, represented by
de Sitter space with Hubble constant $\Hi$, followed by a radiation
dominated universe. The total energy density present is always critical,
$\rt=\rho_{\rm crit}$ . The transition between the two eras can be
approximated as instantaneous, because the physical transition time is much
shorter than the characteristic time for the evolution of the cosmologically
relevant modes, which have $\laph>e^{70}\Hi^{-1}$. The scalar field $\Phi$
is assumed to be non--dominant, therefore it evolves in the background
curved space--time. Back reaction of $\Phi$ on the geometry is neglected,
and gauge ambiguities are eliminated. We consider massless and massive
fields, the latter either stable or decaying into radiation. In this paper
we have not put in a decoherence rate, {\it i.e.} the fields $\Phi$ stay
coherent until they eventually decay. With respect to decoherence properties
the situation is similar to axion models (axions, however, have a non--zero
expectation value) [9]. Since we have not yet included the transfer of the
fluctuations from the resulting massless fields to matter which is
dominating today (cold dark matter and baryons), we give our predictions at
the end of the radiation era.

The state of the quantum field $\Phi$ is the Bunch--Davies state initially,
during inflation: Every observationally relevant mode, which has today
$(\laph)_0<H_0^{-1}$, had at early times in inflation $R/\kph{}^2\to0$ and
is taken to be initially in the Minkowski vacuum state. A physical
discussion of this state can be found in ref. [6].

In section 3 we analyze the contributions of massless fields. We present
several quantities which characterize the energy density fluctuations:
(1) First we derive the energy fluctuations $\dek$ in a single mode $\vk$.
    As long as the mode is super--horizon, {\it i.e.} between the first and
    second horizon crossing, $\dek$ grows linearly with the scale factor
    $a$. This law for massless fields is independent of the cosmological
    era. After the second horizon crossing $\dek\sim a^{-1}$.
(2) Next the equal time two--point correlation function
    $\xi(\ell)=\langle\de(\vx)\de(\vxp)\rangle$ of the energy density
    contrast $\de(\vx)$ is investigated. At the second horizon crossing
    $\xi(\ell)$ is independent of the point separation $\ell$, {\it i.e.} it
    is of Harrison--Zel'dovich type. On super--horizon scales, for
    $\ell\gg\H^{-1}$ where $\H$ is the Hubble parameter, the correlation
    function falls off as $\ell^{-4}$. For $\ell\ll\H^{-1}$ we find a
    logarithmic increase as $(\log\H\ell)^2$ towards smaller scales.
    $\xi(\ell)$ has a cusp at $\ell=2\H^{-1}$, which gives extra strength at
    large scales.
(3) The variance $\ddR$ of the smeared energy density contrast $\de_R$ is
    also scale independent at the second horizon crossing. It grows as
    $(\log\H R)^2$ towards smaller scales for $R\ll\H$, and it decreases as
    $R^{-3}$ on scales $R\gg\H$.
(4) The power spectrum $P(q)$, {\it i.e.} the Fourier transform of
    $\xi(\ell)$, is such that $q^3P(q)$ is scale independent at the second
    horizon crossing, which is again the characteristics of
    Harrison--Zel'dovich type fluctuations. For $q\gg\H$ the quantity
    $q^3P(q)$ is $q$--independent, while on super--horizon scales $(q\ll\H)$
    the power spectrum $P(q)$ itself tends to a constant.
(5) The correlations between the pressure $P$ and energy density $\rho$ are
    computed during the inflationary era. On super--horizon scales (where
    they are independent of the cosmological era) they are related as
    $\de P=-{1\over3}\de\rho$. Using the covariant conservation of the
    energy--momentum tensor this again gives the $a^2$--growth of
    $\sqrt{\xi(\ell)}$ for fixed comoving length.

The magnitude at the second horizon crossing for the relative energy density
fluctuations $\de\rho/\rho$ is of order $(\Hi/\Mp)^2$. Let us assume that
during inflation there were $N$ massless scalar fields fields present and
take for simplicity $N=100$ (in the minimal supersymmetric standard model we
have $N=44$). In order to obtain an amplitude of order $10^{-5}$, we need
$\Hi/\Mp\approx10^{-3}$, {\it i.e.} $\Hi\approx10^{16}$ {\sl GeV} is
required. For the vacuum energy density during inflation,
$\ri:=3\Hi^2\Mp^2/8\pi$, this means $\ri{}^{1/4}\approx2\cdot10^{17}$
{\sl GeV}. This is in conflict with the upper bound of
$\ri{}^{1/4}\lsim5.2\cdot10^{16}$ {\sl GeV} which was derived in [10] from
an estimate of the distortions in the cosmic microwave background radiation
caused by inflationary gravitational waves.

Mass effects for $m\ll\Hi$ are analyzed in section 4 for the energy
fluctuations $\dek$ in a single mode $\vk$. After the super--horizon mode
has become non--relativistic, $\kph\ll m$, we find a growth of
$\dek\sim a^3$, {\it i.e.} it is enhanced over the $\dek\sim a$ law for
massless modes. The growth of $\dek$ stops when $\H$ drops below $m$,
{\it i.e.} when the Compton wavelength $\la_{\rm C}=2\pi/m$ becomes
sub--horizon. If the massive field is stable on the scale of the Hubble
time $\Heq$, $\dek$ is enhanced over the contributions from massless fields
by a $k$--dependent factor at the second horizon crossing. On cosmologically
relevant scales the enhancement factor is $10^{12}$ or more. To compensate
one must take a lower $\Hi$. For an unstable massive field which decays with
a rate $\Gamma$ into radiation the magnitude of $\dek$ is only moderately
enhanced by a factor $\sqrt{m/\Gamma}$, independent of the wave number $k$.

The topic of section 5 is the cosmic variance of $\,\,\or$. In a finite
sample $S$, {\it i.e.} a finite patch of the universe, the spatial average
of the density $\ors=E_S/V_S$ is one measurable number. In a quantum field
theoretic model one can predict the quantum fluctuations of the operator
$E_S$, {\it i.e.} the variance
$(\des)^2=\langle E_S^2\rangle-\langle E_S\rangle^2$, where
$\langle...\rangle$ indicates the (quantum theoretical) ensemble average.
This variance within the ensemble is identified with the variance which
would be observed when measuring $E_S$ in many different patches (of size
$S$) within one universe. For a patch size $\ell_S\simeq30\,h^{-1}$
{\sl Mpc} there are many patches today, and such a variance is accessible to
observations and can be compared to models. For $\ell_S\simeq200\,h^{-1}$
{\sl Mpc} $\,\,\ors$ is (in principle) one measurable number today, but one
might be able to measure its variance in a decade or so when observations
reach much deeper than $200\,h^{-1}$ {\sl Mpc}. For
$\ell_S\simeq H_0{}^{-1}$ observations can give one single number for
$\ors$, but its variance can be inferred from {\sl COBE} measurements under
the assumption of adiabatic fluctuations together with a modest
extrapolation from $\ell=2$ multipoles to $\ell=0$. This gives the cosmic
variance of $\,\,\dors/\rho\simeq10^{-5}$ for $\ell_S\simeq H_0^{-1}$. Our
model gives a contribution to $\dors/\rt$ which drops off as $1/\sqrt{V_S}$
for large sample volumes $V_S$. Finally we relate the experimentally
accessible correlation function for $(\rho(x)-\ors)$, which by definition
must have at least one zero, to the quantum field theoretic correlation
function for $(\rho(x)-\langle\rho\rangle)$, which in our model turns out to
be strictly positive. On small separation scales, $\ell\ll\ell_S$, the
difference between the two correlation functions is $\ell$--independent and
equal to the cosmic variance.

\vskip1truecm

\noindent {\ti 2. Model and Methods}
\vskip0.5truecm

The background is given by a Friedmann--Robertson--Walker (FRW) space--time
with spatially flat sections, {\it i.e.} we take the mean energy density in
the universe to be critical, $\rt=\rho_{\rm crit}$. Using conformal time
$\eta$, we have $ds^2=a(\eta)^2\,(d\eta^2-\vec{dx}{}^2)$. The Hubble
constant during inflation is denoted by $\Hi$. We fix the coordinates in
such a way that at the time $\eta_1$ of the transition from the inflationary
to the radiation dominated stage we have $\eta_1=\Hi^{-1}$ and
$a(\eta_1)=1$. This fixes the scale factor $a(\eta)$ for our cosmological
model as
$$
\alignedat2
a(\eta)\,\,=\,\,\,
 &{1\over(2-\Hi\eta)}\quad&&(\eta\le\Hi^{-1}) \\
a(\eta)\,\,=\,\,\,
 &\Hi\eta            \quad&&(\eta\ge\Hi^{-1}).
\endalignedat
\tag2.1
$$
The Hubble parameter $\H=a^{-2}\partial_\eta a$ at time $\eta$ during the
radiation era reads $\H=a^{-2}\Hi$. The cosmological model (2.1) is
determined by one single model parameter, the Hubble constant $\Hi$ during
inflation.

We investigate the energy fluctuations contributed by a neutral scalar
quantum field $\Phi$ with minimal coupling to gravity. $\Phi$ has no
self--interactions and zero expectation value, $\langle\Phi\rangle=0$. Its
action reads
$$
S_\Phi\,\,=\,\,{1\over 2}\,\int d^4x\,\,\sqrt{-g}\,
            \big(\partial_\mu\Phi\partial^\mu\Phi - m^2\Phi^2\big).
\tag2.2
$$
Variation of the action $S_\Phi$ with respect to the field $\Phi$ gives the
equation of motion (Klein--Gordon equation)
$$
\bigg[\,{d^2\over d\eta^2} + 2{\partial_\eta a\over a}\,{d\over d\eta}
      - {d^2\over d\vx^2} + a^2m^2\,\bigg]\,\Phi\,\,=\,\,0.
\tag2.3
$$
The energy density $\rho$ is always defined to be the one measured by
a comoving observer, who has the 4--velocity $u^\mu\,=\,(a^{-1},0,0,0)$ in
conformal coordinates. Thus we obtain
$$
\rho\,\,=\,\,{1\over2a^2}\,\bigg[\,(\partial_\eta\Phi)^2 +
                          (\partial_{\vx}\Phi)^2 + a^2m^2\Phi^2\,\bigg].
\tag2.4
$$

The energy--energy correlation at equal times is defined by
$$
C(\vx,\vxp)\,:=\,\,\langle\Om|\,\rho(\vx)\rho(\vxp)\,|\Om\rangle
  -\langle\Om|\,\rho(\vx)\,|\Om\rangle\langle\Om|\,\rho(\vxp)\,|\Om\rangle.
\tag2.5
$$
The quantum state $|\Om\rangle$ is initially (during inflation) the
Bunch--Davies state: Every observationally relevant mode, which has today
$(\laph)_0<H_0^{-1}$, had at early times in inflation $R/\kph{}^2\to0$ and
is taken to be initially in the Minkowski vacuum state. Since $|\Om\rangle$
is translation invariant, $C$ depends only on the physical distance
$\ell=a\,|\vx-\vxp|$ of the two points.

We use normal ordering to define the renormalized energy density operator
\linebreak
$\rho(x):=$N$[\rho(x)]$. The key observation [6] is that different choices
for the normal ordering prescription give identical results for the energy
autocorrelation function $C(\vx,\vxp)$. This is so because two normal
orderings N$[\rho]$ differ by a c--number, which drops out in the connected
two--point function $C(\vx,\vxp)$.

The mode functions in a spatially flat FRW space--time are eigenfunctions of
the comoving wave vector $\vk$. Because of the translational invariance of
the gravitational field in 3--space different $\vk$'s decouple, and $\vk$ is
a conserved quantity. We make the ansatz for the modes of fixed $\vk$
$$
\varphi_k(\eta,\vx)\,\,=:\,u_k(\eta)\,e^{i\vk\vx}.
\tag2.6
$$
The modes $u_k(\eta)$ are normalized solutions of the dynamics throughout
time, {\it i.e.} they are solutions of the Klein--Gordon equation (2.3) in
both eras of our cosmological model, and $u_k$ and $\partial_\eta u_k$ are
continuous at the transition. We call $\uki$ those modes which have the time
dependence $e^{-ik\eta}$ at early times, as $k\eta\to-\infty$ in inflation.
The $\uki(\eta)$ are those evolving modes which approached Minkowski single
particle waves at early times (when their physical wavelengths were much
smaller than the Hubble radius).

Since the results are independent of the normal ordering, we can make the
most convenient choice, the one adapted to the quantum state
$|\Omega\rangle$. We use the mode expansion of the field operator
$\Phi(\eta,\vx)$ as
$$
\Phi(\eta,\vx)=\int\,{d^3k\over (2\pi)^3}
\left[\,\aki      \uki(\eta)\,  e^{i\vk\vx} +
        \aki{}^\dag \uki(\eta)^*\,e^{-i\vk\vx}\,\right].
\tag2.7
$$
The state $|\Om\rangle$ of the quantum field is annihilated by all the
operators $\aki$,
$$
\aki|\Om\rangle = 0,\quad \hbox{for all}\,\,k.
\tag2.8
$$

A key tool of ref. [6] was that the computation of the correlation function
$C(\vx,\vxp)$ in a given quantum state $|\Om\rangle$ can be reduced to the
computation of the Wightman function of this state,
$$
W(x,x')\,:=\,\,\langle\Om|\,\Phi(x)\,\Phi(x')\,|\Om\rangle
\,\,=\,\,\int {d^3k\over(2\pi)^3}\,\,\uki(\eta)\uki(\eta')^*\,
                                                         e^{i\vk(\vx-\vxp)}.
\tag2.9
$$
In the massless case, the two--point correlation function $C(\vx,\vxp)$ is
given as
$$
C(\ell)\,\,=\,\,
 {1\over2a^4}\,\sum_{\al,\be=0}^3\,\,
         \big[W(x,x')_{,\al\be'}\big]^2\,\,\bigg|_{\dsize{\eta=\eta'}},
\tag2.10
$$
where primed derivatives act exclusively on the primed coordinates.

The inflationary era was treated in ref. [6]. In particular, for $m=0$ the
modes $\uki$ have the functional form
$$
\uki(\eta)\,\,=\,\,
   {\Hi\over\sqrt{2k^3}}\,(i-k\tilde\eta)\,e^{-ik(\tilde\eta+\eta_1)}
                                           \quad(m=0,\,\,\eta\le\Hi^{-1}),
\tag2.11
$$
where $\tilde\eta=\eta-2\Hi^{-1}$, {\it i.e.} they have the time dependence
of a Hankel function, \linebreak
$\uki\sim\tilde\eta^{3/2}\,$H$_{3/2}{}^{(2)}(k\tilde\eta)$. The two--point
correlation function $C$ was found as
$$
C(\ell)\,\,=\,\,{1\over(2\pi)^4}\,
 \bigg[\,{24\over\ell^8}+{14\Hi^2\over\ell^6}+{3\Hi^4\over2\ell^4}\,\bigg]
 \quad (m=0,\,\,\eta\le\Hi^{-1}).
\tag2.12
$$

The energy density contrast contributed by the scalar field under
consideration is
$$
\de(x)\,\,=\,\,{1\over\rt}\big(\rho(x)-\langle\rho\rangle\big),
\tag2.13
$$
where $\rt = 3\H^2\Mp^2/8\pi$. The two--point function of $\de(x)$,
$$
\xi(\ell)\,\,=\,\,\langle\Om|\,\de(\vx)\de(\vxp)\,|\Om\rangle,
\tag2.14
$$
is a measure of the relative energy density fluctuations. The two--point
functions (2.5) and (2.14) are related as $\xi(\ell)\equiv C(\ell)/\rt^2$.

\vskip1truecm

\noindent{\ti 3. The Energy Fluctuations for Massless Fields}
\vskip0.5truecm
\noindent {\bf3.1.} {\bfsl The evolved modes} \br\indent
After the transition to the radiation dominated era, at times
$\eta\ge\eta_1=\Hi^{-1}$, the Klein--Gordon equation (2.3) reads
$$
\bigg[\,{d^2\over d\eta^2} + {2\over\eta}\,{d\over d\eta}
      + k^2 +m^2\Hi^2\eta^2\,\bigg]\,u_k(\eta)\,\,=\,\,0.
\tag3.1
$$
We choose as fundamental solutions for $m=0$
$$
v_k(\eta)\,\,=\,\,{1\over\sqrt{2k}\,\Hi\eta}\,e^{-ik(\eta-\eta_1)}
\tag3.2
$$
and its complex conjugate $v_k^*$. They have the time dependence
$\eta^{-1/2}\,$H$_{1/2}^{(1,2)}(k\eta)$. The mode functions $\uki$, which
correspond to the initial Bunch--Davies state, are linear combinations of
the two fundamental solutions $v_k$ and $v_k^*$,
$$
\uki(\eta)\,\,=\,\,\al_k\,v_k(\eta) + \be_k\,v_k(\eta)^*.
\tag3.3
$$
The $\al_k$ and $\be_k$ in eq. (3.3) are called Bogolubov coefficients.
They are fixed by requiring a continuous transition of the mode functions
$\uki(\eta)$, given by eqs. (2.11) and (3.3), and their first derivative at
time $\eta=\eta_1=\Hi^{-1}$. We obtain
$$
\al_k\,\,=\,\,1+{i\Hi\over k}-{\Hi^2\over2k^2}\qquad\hbox{and}\qquad
\be_k\,\,=\,\,{\Hi^2\over2k^2}.
\tag3.4
$$
This satisfies the general condition for Bogolubov coefficients
$|\al_k|^2-|\be_k|^2=1$. We get the functional form of the modes $\uki$
during the radiation era as
$$
\uki(\eta)\,\,=\,\,{1\over\sqrt{2k}}\,{1\over \Hi\eta}\,
 \bigg[\,\bigg(1+i{\Hi\over k}\bigg)\,e^{-ik(\eta-\eta_1)}\,+\,
                          i{\Hi^2\over k^2}\,\sin k(\eta-\eta_1)\,\bigg].
\tag3.5
$$
There are two types of modes, those which never went beyond the Hubble
radius, {\it i.e.} $\laph<\H^{-1}$ at all times, and those modes which spent
some time with $\laph>\H^{-1}$. The separation point is $k=\Hi$. From now on
we look exclusively at the cosmologically relevant modes, which spent a long
time outside the Hubble radius, {\it i.e.} at modes with $k\ll\Hi$,
$$
\uki(\eta)\,\,=\,\,
 {i\over\sqrt{2k}}\,{\Hi\over k}\,{\sin k\eta\over k\eta}
                                          \quad (\hbox{for}\,\,k\ll\Hi).
\tag3.6
$$
Since $k\eta=\kph/\H$ we see that well before the second horizon crossing,
$\kph\ll\H$, $\uki(\eta)$ is time independent. After the second horizon
crossing it oscillates in time.

\noindent{\bf 3.2.} {\bfsl Energy fluctuations in one mode}\br\indent
We start our investigation of the energy fluctuations by looking at one
single mode $\vk$. The mode decomposition (Fourier transform) will be
defined in a comoving cube $L^3$ with periodic boundary conditions. We
compute the fluctuations of the total measured energy $\ek$ within the
periodicity cube, contributed by the given mode $\vk$,
$$
(\dek)^2\,:=
        \,\,\langle\Om|\,\ek^2\,|\Om\rangle-\langle\Om|\,\ek\,|\Om\rangle^2.
\tag3.7
$$
The energy of a single--particle quantum measured at a given time is
$$
\emk\,\,=\,\,\sqrt{\kph^2+m^2},
\tag3.8
$$
and the fluctuations of the number of $\vk$--quanta at a given time are
$\dnk=\dek/\emk$.

The operator $E_V$ of the total energy in the physical volume $V=(aL)^3$
relates the energies $\ek$ to the energy density $\rho$, eq. (2.4),
$$
E_V\,\,=\,\,a^3\,\int_{\dsize L^3} d^3x\,\,\rho(x)\,\,=\,\,\sum_{\vk}\,\ek.
\tag3.9
$$
Using the discrete version of the mode decomposition (2.7) of the field
operator $\Phi(x)$ we find the operator $\ek$ in terms of bilinears of the
annihilation and creation operators $\aki$ and $\aki{}^\dag$, accompanied by
$\uki$ and $\uki^*$ with their derivatives. The fluctuations of $\ek$ are
given by $|\langle\Om|\ek|\vk,-\vk\rangle|^2$, and we obtain
$$
(\dek)^2\,\,=\,\,
  {a^6\over4}\,\bigg[\,{(W_{k,\eta\eta'})^2\over a^4} +
           2\,{\hbox{Re}\,(W_{k,\eta})^2\over a^2}\,\emk^2 +
                (W_k)^2\,\emk^4\,\bigg]\quad\bigg|_{\dsize\eta=\eta'},
\tag3.10a
$$
where
$$
W_k\,:=\,\,\uki(\eta)\uki(\eta')^*
\tag3.10b
$$
is Fourier transform of the Wightman function (2.9).

During inflation the mode functions $\uki$ are given by eq. (2.11), and we
obtain
$$
\dek\,\,=\,\,{\Hi^2\over4\kph}\,\sqrt{1+4{\kph^2\over\Hi^2}}.
\tag3.11
$$
In the radiation era the mode functions $\uki$ are given by eq. (3.6), and
we obtain the energy fluctuations
$$
\dek\,\,=\,\,{\Hi^2\H^2\over4\kph^3}\,
     \bigg(\,1 - {\sin2\x\over\x} + {\sin^2\x\over\x^2}\,\bigg),
     \quad {\x}\,=\,{\kph\over\H}\,=\,k\eta.
\tag3.12
$$
For very small and very large wave numbers compared to the Hubble parameter
this simplifies to
$$
\dek\,\,\simeq\quad
\left\{
 \alignedat2
    &\,\,{\Hi^2\over4\kph}      \quad&&(\hbox{for}\,\,\kph\ll\H) \\
    &\,\,{\Hi^2\H^2\over4\kph^3}\quad&&(\hbox{for}\,\,\H\ll\kph\ll{\Hi\over
a}).
 \endalignedat
\right.
\tag3.13
$$
Let us summarize the time evolution of $\dek$ for a fixed comoving wave
number $k\ll\Hi$, see {\bfsl fig. 3.1}:
(1) Before the first horizon crossing early in the inflationary era, $\dek$
    is constant, and $\dnk\to0$ for $\kph\gg\Hi$.
(2) At the first horizon crossing $\dek=\Cal{O}(\Hi)$ and $\dnk=\Cal{O}(1)$.
(3) Between the first and the second horizon crossing, when $\kph\ll\Hi$ and
    $\kph\ll\H$, the fluctuations $\dek$ grow in time linearly with the
    scale factor $a$, and $\dnk$ grows as $a^2$. This law remains valid as
    long as the mode is super--horizon sized, independent of the
    cosmological era.
(4) After the second horizon crossing, for $\kph\gg\H$ during the radiation
    dominated era, the fluctuations $\dek$ decrease in time $\sim a^{-1}$.
    The fluctuations $\dnk$ do not evolve in time any more. The total growth
    factor for $\dnk$ between the first and second horizon crossing is
    $\Hi^4/k^4$, a factor $\Hi^2/k^2$ between the first horizon crossing and
    the end of inflation, another factor $\Hi^2/k^2$ during the radiation
    era up to the second horizon crossing.

At fixed time in the radiation era we find $\dek\sim k^{-3}$ for
sub--horizon modes, $\kph\gg\H$. For super--horizon modes, with $\kph\ll\H$,
we have $\dek\sim k^{-1}$. The two asymptotes meet at $\kph=\H$.

\noindent{\bf 3.3.} {\bfsl Two--point correlation function}\br\indent
We now sum up the contributions of all the $\vk$--modes to the energy
density fluctuations. To do this we approximate the mode functions
$\uki(\eta)$ again by eq. (3.6) alone. Thus we will obtain an approximation
for $\xi(\ell)$ which is valid only on length scales $\ell\gg a\Hi^{-1}$.
But it will fail for the short--distance behaviour of $\xi(\ell)$, on
scales $\ell$ which never went beyond the Hubble radius. However, the
short--distance behaviour of $\xi(\ell)$ is universal, independent of
quantum state and external curved space--time. This term was discussed in
ref. [6], and it can be read off eq. (2.12).

The integral (2.9) for $W(x,x')$ is logarithmically divergent in the
infrared, but the divergent term is independent of the coordinates of the
two points $x$ and $x'$ and therefore drops out when computing $\xi(\ell)$.
We insert (3.6) into (2.9) and obtain for the infrared finite terms of the
Wightman function
$$
\aligned
W(x,x')\,\,=\,\,
{\Hi^2\over(2\pi)^2}{1\over24\eta\eta'\vt x}\,
\bigg[&\,\Delta_{-+}^3\,\log |\Hi\Delta_{-+}|\,\,+\,\,
         \Delta_{+-}^3\,\log |\Hi\Delta_{+-}| \\
   &-\,\,\Delta_{--}^3\,\log |\Hi\Delta_{--}|\,\,-\,\,
         \Delta_{++}^3\,\log |\Hi\Delta_{++}|\,\,\bigg].
\endaligned
\tag3.14
$$
We have used the notation $\vt x:=|\vx-\vxp|$ and
$\Delta_{\pm\pm}:=\,\,\vt x\pm\eta\pm\eta'$.

First we investigate the correlation function $\xi(\ell)$ on sub--horizon
scales, \linebreak $a\Hi^{-1}\ll\ell\ll\H^{-1}$. We obtain
$$
\xi(\ell)\,\,\simeq\,\,
   {1\over6}\,\bigg({2\Hi^2\over3\pi\Mp^2}\bigg)^2\,
                                          \bigg(\log{\H\ell\over2}\bigg)^2
 \quad (a\Hi^{-1}\ll\ell\ll\H^{-1}),
\tag3.15
$$
a plateau--like structure on scales well inside the Hubble radius. Here
$\xi(\ell)$ increases logarithmically towards the smaller scales. On
super--horizon scales, $\H\ell\gg1$, we find
$$
\xi(\ell)\,\,\simeq\,\,
   {3\over2}\,\bigg({2\Hi^2\over3\pi\Mp^2}\bigg)^2\,{1\over(\H\ell)^4}
 \quad (\ell\gg\H^{-1}),
\tag3.16
$$
a power law decay. In the range of intermediate length scales
$\ell\approx\H^{-1}$ we need the exact expression for the correlation
function $\xi(\ell)$,
$$
\aligned
\xi(\ell)\,\,=\,\,
  \bigg({2\Hi^2\over3\pi\Mp^2}\bigg)^2\,
  \bigg[&\,\bigg({1\over48\y^4}+{1\over144\y^2}+{1\over72} \bigg)\,+\,
        \Cal{L}_1\,  \bigg(-{1\over48\y^5}+{1\over27\y^3}-{5\over144\y}\bigg)\\
&+\,    \Cal{L}_2\,  \bigg({5\over72}-{\y^2\over36} \bigg)\,
 +\,    \Cal{L}_1^2\,\bigg({1\over192\y^6}-{5\over144\y^4}+{17\over576\y^2}
                                                         +{1\over64}\bigg)\\
&\hskip1.5truecm+\,
 \Cal{L}_1\Cal{L}_2\,\bigg(-{1\over18\y}-{\y\over36}\bigg)\,
 +\,    \Cal{L}_2^2\,\bigg({1\over24}-{\y^2\over72}+{\y^4\over72}\bigg)
                                                                   \,\bigg].
\endaligned
\tag3.17
$$
We have used the dimensionless variable $\y={1\over2}\H\ell$, further
$\Cal{L}_1:=\log|(\y+1)/(\y-1)|$ and $\Cal{L}_2:=\log|\y^2/(\y^2-1)|$.
The function $\xi(\ell)$ is strictly positive for all $\ell$, see
{\bfsl fig. 3.2}.

The correlation function $\xi(\ell)$ is finite for all $\ell>0$, but it is
not differentiable at the point $\ell=2\H^{-1}$. This can already be seen in
the Wightman function $W(x,x')$, eq. (3.14). The term involving
$\Delta_{--}$ has a pole in its third derivatives. This cusp singularity has
its origin in the instantaneous transition from inflation to the radiation
era. We can mimic a finite duration by introducing an exponential high
frequency cut--off $e^{-kT}$ into the Bogolubov coefficient $\be_k$ of eq.
(3.4), where $T$ is the fastest comoving time scale involved in the phase
transition (see the discussion in ref. [11]). However, since in any case
the time scale $T$ is at most one or two orders of magnitude away from the
Hubble time during inflation $\Hi^{-1}$, the smoothing of the cusp occurs on
an extremely small scale at the time of matter and radiation equality. The
effect is invisible on the plot.

The correlation function $\xi(\ell)$ depends on $\ell$ only via $\H\ell$,
{\it i.e.} it is time independent if evaluated at fixed $\ell/\ell_{\H}$,
where $\ell_{\H}$ is the Hubble radius. At the second horizon crossing, when
$\H\ell={\rm fixed}=\Cal{O}(1)$, the correlation is $\ell$--independent and
therefore of Harrison--Zel'dovich type. The magnitude is of order
$(\Hi/\Mp)^4$. For instance, we obtain
$\sqrt{\xi(\ell=\H^{-1})}=0.098(\Hi/\Mp)^2$.

\noindent{\bf 3.4.} {\bfsl Variance of the smeared energy density
contrast}\br\indent
We introduce the spatially smeared energy density contrast
$$
\de_R\,:=\,\,\int d^3x\,\,\de(\vx)\,W_R(\vx).
\tag3.18
$$
For convenience, we take a Gaussian window function $W_R$ of scale $R$,
$$
W_R(\vx)\,\,=\,\,{1\over(\sqrt{2\pi}\,R)^3}\,e^{\tsize-{x^2\over2R^2}}.
\tag3.19
$$
The variance is given by
$\ddR=\int d^3x\int d^3y\,W_R(\vx)W_R(\vy)\,\xi(\vx,\vy)$.
Since the equal time two--point function $\xi$ depends only on the physical
separation $\ell$ of the two points, one integration is trivial. We obtain
$$
\ddR\,\,=\,\,{1\over2\sqrt{\pi}R^3}\,\int_0^\infty {d\ell\over\ell}\,
                        \ell^3\,\,e^{\tsize-{\ell^2\over4R^2}}\,\,\xi(\ell).
\tag3.20
$$

We first evaluate $\ddR$ for sub--horizon smearing scales, $\H R\ll1$. For
sub--horizon separations, $\H\ell\ll1$, the two--point function $\xi(\ell)$
varies only logarithmically, see eq. (3.15). Therefore the integral (3.20)
is strongly peaked at $\ell=\sqrt{6}R$, and we obtain
$$
\ddR\,\,\simeq\,\,\sqrt{3\over2\pi e^3}\,\bigg({2\over3\pi}\bigg)^2\,
                       \bigg({\Hi\over\Mp}\bigg)^4\,\,\big(\log\H R\big)^2
\quad (\H R\ll1).
\tag3.21
$$
For super--horizon smearing scales, $\H R\gg1$, the exponential in eq.
(3.20) can be dropped and
$$
\ddR\,\,\simeq\,\,{88\sqrt{\pi}\over1890}\,\bigg({\Hi\over\Mp}\bigg)^4\,
                                           {1\over(\H R)^3}\quad(\H R\gg1).
\tag3.22
$$
The variance $\ddR$ is plotted in {\bfsl fig. 3.3}. For $\H R\gg1$ we have
an $R^{-3}$ fall--off, in contrast to the $\ell^{-4}$ super horizon
fall--off for $\xi(\ell)$. The reason for this is simple: Because
$\xi(\ell)$ falls off rapidly for $\ell\gg\H^{-1}$, the integral (3.20) is
strongly peaked at $\ell\approx\H^{-1}$, independent of the scale $R$, if
$R\gg\H^{-1}$. In this limit the $R$--dependence is entirely due to the
prefactor in eq. (3.20), which reflects the fact that $\ddR$ involves three
non--trivial integrations.

\noindent{\bf 3.5.} {\bfsl Power spectrum}\br\indent
Up to now we have concentrated on the two--point correlation function in
$x$--space. For many purposes it is useful to have the Fourier transform
of $C(\ell)$,
$$
C(q)\,\,=\,\,{1\over a^3}\,\int d^3\ell\,\,C(\ell)\,e^{-i\vq\vec{\ell}/a}
\tag3.23
$$
(or correspondingly the power spectrum $\xi(q)=C(q)/\rt^2$), where
$a^{-1}\ell$ is the comoving separation. By inserting eq. (2.10) we obtain
in terms of $u_k\equiv\uki(\eta)$
$$
\aligned
C(q)\,\,=\,\,{1\over2a^4}\,\int {d^3k\over(2\pi)^3}\,
 \bigg\{\,|\partial_\eta u_k|^2\,|\partial_\eta u_{|\vq-\vk|}|^2
  +(\vq\,\vk-\vk\,^2)^2\,|u_k|^2\,|u_{|\vq-\vk|}&|^2 \\
 -\,2(\vq\,\vk-\vk\,^2)\,\hbox{Re}\,\big[u_k(\partial_\eta u_k)^*\,
            u_{|\vq-\vk|}(\partial_\eta u_{|\vq-\vk|})^*\big]     &\,\bigg\}.
\endaligned
\tag3.24
$$
Note that the Fourier transform $C(q)$ of the two--point correlation
function $C(\ell)$ involves convolutions of the Wightman function in
Fourier space,
$W_{\vk}:=\langle\Om|\,\Phi(\vk)\Phi(-\vk)\,|\Om\rangle=|\uki(\eta)|^2$
corresponding to the squares occurring in eq. (2.10). The mode functions
$\uki$ are given by eq. (3.6). The resulting power spectrum $C(q)$ is
plotted in {\bfsl fig. 3.4}. In the limit $q\to0$ the correlation function
$C(q)$ tends to the integral of $C(\ell)$ over all space, see eq. (3.23).
Since $C(\ell)$ is positive everywhere, $C(q)$ must tend to a positive
constant for $q\to0$. We obtain
$$
C(q)\,\,\simeq\,\,{1\over a^3}\,\int d^3\ell\,\,C(\ell)\,\,=\,\,
     {11\over420\pi}\,{\Hi^4\H\over a^3}\quad\hbox{for}\,\,\qph\ll\H.
\tag3.25
$$
We write $C(q=0)$ in terms of single mode contributions. From eqs. (3.24)
and (3.10) we obtain in the continuum limit $L\to\infty$
$$
C(q=0)\,\,=\,\,\int {d^3k\over(2\pi)^3} \,\,(\drk)^2,
\tag3.26
$$
where $\drk:=V^{-1}\,\dek$ are the fluctuations in the mean energy density
$\rk=V^{-1}\ek$ in the physical volume $V=(aL)^3$ contributed by the mode
$\vk$. $C(q=0)$ is a sum of squares, and it cannot vanish unless $\dek=0$
for all modes $\vk$. In the standard treatment of inflaton fluctuations only
the interference terms between the fluctuations $\de\Phi_{\vk}$ and the
homogeneous background are kept. The interference terms cannot contribute to
the sum of squares on the right hand side of eq. (3.26), and therefore
$C(q)_{\rm inflaton}\to0$ for $q\to0$.

In the sub--horizon wavelength region, $\qph\gg\H$, we find numerically
$$
C(q)\,\,\simeq\,\,
  {5\over32\pi^2}\,{\Hi^4\H\over a^3}\,\bigg({\H\over\qph}\bigg)^3
                              \quad\hbox{for}\,\,\H\ll\qph\ll{\Hi\over a}.
\tag3.27
$$
When evaluated at the second horizon crossing, where
$\qph/\H={\rm fixed}=\Cal{O}(1)$, the quantity $q^3\xi(q)$ is scale
independent. The energy density fluctuations are of Harrison--Zel'dovich
type. For instance, we find numerically
$q^3\xi(\qph=\H)\simeq{1\over6}\,(\Hi/\Mp)^4$.

The relative energy density fluctuations for a fixed comoving $\vq$ grow in
time as $\xi(q)\sim a^3$ before the second horizon crossing. This is in
contrast to the energy density fluctuations in the inflaton scenario which
are also of Harrison--Zel'dovich type ({\it i.e.} scale independent at the
second horizon crossing), but which show a super--horizon growth as
$\xi(q)_{\rm inflaton}\sim a^4$. Therefore $C(q)_{\rm inflaton}\sim q$ for
$\qph\ll\H$.

\noindent{\bf3.6.} {\bfsl Two--point correlations of the
                          energy--momentum--stress tensor}\br\indent
We investigate the equal--time connected two--point correlation functions
of the energy--momentum--stress tensor,
$$
C_{\hat a\hat b\hat c\hat d}(\vec\ell\,)\,\,=\,\,
 \langle\Om|\,T_{\hat a\hat b}(\vx)\,
             T_{\hat c\hat d}(\vxp)\,|\Om\rangle -
        \langle\Om|\,T_{\hat a\hat b}\,|\Om\rangle
                  \langle\Om|\,T_{\hat c\hat d}\,|\Om\rangle,
\tag3.28
$$
where the hatted indices refer to the local orthonormal frame of a comoving
observer. For simplicity, we restrict ourselves to the inflationary de
Sitter era. These correlation functions depend on the separation vector
$\vec\ell=(\vx-\vxp)_{\rm phys}$. The computation goes along the same line
as outlined in section 2, and the correlation functions can be expressed in
terms of the Wightman function $W(x,x')$ as
$$
\aligned
C_{\hat a\hat b\hat c\hat d}(\vec\ell\,)\,\,=\,\,
 W_{\lower2pt\hbox{${\ssize,\hat a\hat c'}$}}\,W_{,\hat b\hat d'} +
 W_{,\hat a\hat d'}\,W_{,\hat b\hat c'} -
 \eta_{\hat a\hat b}\,
     W_{\lower2pt\hbox{${\ssize,\hat z\hat c'}$}}\,&
                          W_,{}^{\hat z}{}_{\hat d'} -
 \eta_{\hat c\hat d}\,
     W_{\lower2pt\hbox{${\ssize,\hat a\hat z'}$}}\,
                                           W_{,\hat b}{}^{\hat z'} \\
 &+ {1\over2}\,\eta_{\hat a\hat b}\,\eta_{\hat c\hat d}\,
        W_{\lower2pt\hbox{${\ssize,\hat y\hat z'}$}}\,
                                        W_,{}^{\hat y\hat z'}.
\endaligned
\tag3.29
$$
For $a=b=c=d=0$ we recover eq. (2.10). The tensor structure of
$C_{\hat a\hat b\hat c\hat d}$ can be built using $\eta_{\hat a\hat b}$ and
$\eps_{\hat a}=\ell_{\hat a}/\ell$. To simplify we average over the
directions $\vec\eps$, and therefore the indices of
$C_{\hat a\hat b\hat c\hat d}$ must be pairwise equal. The possibilities are
$\rho\rho$--, $\rho P$--, $PP$--, and $\vec{S}\cdot\vec{S}$--correlations,
where $P={1\over3}(T_{\hat1\hat1}+T_{\hat2\hat2}+T_{\hat3\hat3})$ is the
isotropic pressure, and $\vec{S}$ is the energy flow density (or momentum
density), $S^{\hat a}=T^{\hat0\hat a}$.

The computation goes along the same lines as outlined in section 2. For
super--horizon separations, $\ell\gg\Hi^{-1}$, the energy density and
pressure correlations are all of the order $\Hi^4\ell^{-4}$ with relative
magnitudes
$$
C_{PP}(\ell)\,\,=\,\,{1\over9}\,C_{\rho\rho}(\ell),\quad
        C_{\rho P}(\ell)\,\,=\,\,-\,{1\over3}\,C_{\rho\rho}(\ell)
                                             \quad\hbox{for}\,\,\Hi\ell\gg1,
\tag3.30
$$
with $C_{\rho\rho}$ given in eq. (2.12). The ratios correspond to an
effective equation of state
$$
\de P\,\,=\,\,-{1\over3}\,\de\rho.
\tag3.31
$$

Using this effective equation of state and the conservation law
$T^{\mu\nu}{}_{;\nu}=0$ we can derive the growth of the super--horizon
perturbations. In a small comoving volume $V$ of the external FRW space the
conservation law is equivalent to d$U=-P$d$V$ and to \linebreak
d$\rho=-3(\rho+P)($d$a/a)$. Writing $\rho-\langle\rho\rangle=\de\rho$ we
obtain
$$
{\rm d}(\de\rho)\,\,=\,\,-3(\de\rho + \de P)\,{{\rm d}a\over a}.
\tag3.32
$$
The equation of state (3.31) yields $\de\rho\sim a^{-2}$, the
super--horizon growth factor which we have found earlier using a different
method, see eq. (3.16).

For the autocorrelation function of the energy flow density $\vec{S}$ we
find on super--horizon scales
$$
C_{SS}(\ell)\,\,\simeq\,\,-\,{1\over(2\pi)^4}\,{6\Hi^2\over\ell^6}
                                             \quad\hbox{for}\,\,\Hi\ell\gg1.
\tag3.33
$$
The correlation function of the flow density decreases much faster with
$\ell$ than the correlations of the densities $\rho$ and $P$.

In ref. [12] correlations of the energy--momentum--stress tensor were
computed which are not averaged over directions. In this case one must
distinguish pressures and energy flows longitudinal and transversal to
$\vec{\ell}$.

\vskip1truecm

\noindent{\ti 4. Mass Effects}
\vskip0.5truecm
\noindent{\bf4.1.} {\bfsl Stable massive fields}\br\indent
In this section we investigate the time evolution of $\dek$ for a mode $\vk$
of a "light" massive scalar with $m\ll\Hi$. As in the massless case of
section 3, we consider only modes that spend a considerable amount of time
outside the horizon, {\it i.e.} with $\kph\ll\Hi$ at the end of inflation.
As long as $m\ll\kph$ the mode is ultra--relativistic and behaves as if it
was massless. It becomes non--relativistic ($\kph\approx m$) only after the
first horizon crossing.

The mass $m$ introduces a second length scale, the Compton wavelength
$m^{-1}$, in addition to the Hubble length $\H^{-1}$ of the external
background. Recall that in the massless case everything depends on the
relation of the two characteristic scales \linebreak
$\emk=\kph$ and $\H$. The horizon crossings occur when $\kph\approx\H$. If
$m\neq0$ the ratio $\emk/\H$, where $\emk{}^2=\kph^2+m^2$, is dynamically
relevant for the horizon crossing and not $\kph/\H$. In addition, the ratio
$\kph/m$ determines the transition from the relativistic to the
non--relativistic regime. A typical scenario is plotted in {\bfsl fig. 4.1}.

During inflation the massive modes are given in terms of Hankel functions
[6], $\uki(\eta)={1\over2}\sqrt{\pi\tilde\eta^3}\Hi\,
                                           {\rm H}_\nu{}^{(2)}(k\tilde\eta)$
of order $\nu^2={9\over4}-m^2\Hi^{-2}$, where $\tilde\eta=\eta-2\Hi^{-1}$.
Using eq. (3.10) we obtain for super--horizon modes
$$
\dek\,\,\simeq\,\,{\Hi^2\over4}\,{\emk^2\over\kph^3}
                    \quad({\rm for}\,\,\emk\ll\Hi\,\,{\rm and}\,\,m\ll\Hi).
\tag4.1
$$
Here we consider only masses $(m/\Hi)^2\ll{1\over70}$. Since cosmologically
relevant modes have $\kph/\Hi=\Cal{O}(e^{-70})$ at the end of inflation,
factors $(\kph/\Hi)^{m^2/\Hi^2}$ are irrelevant in eq. (4.1). As long as
the mode is relativistic, {\it i.e.} when $\emk\simeq\kph$, the fluctuations
$\dek$ grow in time linearly with the scale factor $a$. In the
non--relativistic regime the growth is accelerated to $\dek\sim a^3$, since
the physical energy $\emk\simeq m$ is no longer redshifted.

During the radiation dominated era the Klein--Gordon equation is given by
eq. (3.1). For a highly non--relativistic mode, with $\kph\ll m$, we can
drop the $k$--dependence. We choose as fundamental solutions
$$
v_k(\eta)\,\,\simeq\,\,\sqrt{{\pi\over32\Hi{}^2\eta}}\,\,
  \hbox{H}_{{\tsize{1\over4}}}{}^{(2)}\big({\Hi m\over2}\,\eta^2\big)
                                                          \quad(\kph\ll m)
\tag4.2
$$
and its complex conjugate, $v_k(\eta)^*$. The mode functions $\uki$ are
linear combinations of $v_k$ and $v_k^*$, constructed in such a way that
$\uki$ and its first derivative are continuous at the time $\eta=\Hi^{-1}$
of the transition from inflation to the radiation era. We obtain
$$
\uki(\eta)\,\,\simeq\,\,{i\Gamma({\tsize{5\over4}})\over\sqrt{8\Hi^2\eta}}\,
 \bigg({\Hi\over m}\bigg)^{{\tsize{1\over4}}}\,
           \bigg({k\over2\Hi}\bigg)^{\tsize -{3\over2}}\,
           \hbox{J}_{{\tsize{1\over4}}}\big({\Hi m\over2}\,\eta^2\big)
                                                          \quad(\kph\ll m).
\tag4.3
$$
In the modes $\uki$ and in the following expressions we neglect powers of
$(k/\Hi)^{m^2/\Hi^2}$. The energy fluctuations (3.10) are
$$
\dek\,\,=\,\,\Gamma({\tsize{5\over4}})^2{\Hi^2m^2\over2\kph^3}\,
  \sqrt{{\H\over m}}\,\bigg(\,
     {\rm J}_{{\tsize{5\over4}}}\big({m\over2\H}\big)^2 +
     {\rm J}_{{\tsize{1\over4}}}\big({m\over2\H}\big)^2 \,\bigg)
                                                          \quad(\kph\ll m).
\tag4.4
$$
We distinguish two asymptotic limits
$$
\dek\,\,\simeq\quad
\left\{\quad
  \alignedat2
    &{\Hi^2m^2\over4\kph^3} \quad&&(\emk\simeq m\ll\H) \\
    &{\Gamma({\tsize{1\over4}})^2\over8\pi}\,{\sqrt{\Hi^7m}\over k^3}
                            \quad&&(\emk\simeq m\gg\H).
  \endalignedat
\right.
\tag4.5
$$
As long as the mode is super--horizon, $\dek$ goes on growing in time as
$a^3$ independent of the cosmological era. After the second horizon
crossing, when $\emk\simeq m\gg\H$, the fluctuations $\dek$ do not evolve in
time any more, $\dek$ is constant in time. The time evolution of $\dek$ is
plotted in {\bfsl fig. 4.2}.

The most interesting effect of the mass $m$ is that the energy density
fluctuations $\dek$ grow faster in time as soon as the super--horizon mode
becomes non--relativistic, when $\kph\approx m$ and $\emk\ll\H$. As a
consequence, the fluctuations are much bigger for a massive than for a
massless scalar field. It can easily be read off figs. 3.1 and 4.2 that the
ratio at the time when $\kph=\H$ ({\it i.e.} at the second horizon crossing
for $\kph$) is
$$
{(\dek)_{m\neq0}\over(\dek)_{m=0}}\,\,=\,\,{\sqrt{\Hi m}\over k}
                                             \quad(\hbox{for}\,\,\kph=\H).
\tag4.6
$$
The ratio depends on the comoving wave number $k$, which equals $\kph$ at
the end of inflation. The magnitude of this ratio for modes with
$\kpheq=\Heq$, {\it i.e.} with $\laph\simeq160\,h^{-1}$ {\sl Mpc} today, is
$$
{(\dek)_{m\neq0}\over(\dek)_{m=0}}\,\,=\,\,\sqrt{{m\over\Heq}}
                     \,\,\simeq\,\,{\sqrt{m\Mp}\over1\,\,\hbox{\sl eV}}
                     \,\,\simeq\,\,{\sqrt{m\over10^{-28}\,\,\hbox{\sl eV}}}
                                \quad(\hbox{for}\,\,\kpheq=\Heq).
\tag4.7
$$
Note that the dependence on $\Hi$ has dropped out from this ratio since for
fixed $\kpheq$ the physical wave number at the end of inflation varies as
$\sqrt{\Hi}$. The massive fluctuations are enhanced over the massless
fluctuations by an enormous factor for physically reasonable non--zero
masses. For instance with $m\gsim10^{-3}$ {\sl eV} the enhancement factor is
larger than $10^{12}$. Since massless modes with horizon crossing at
equality have $(\dek)_{m=0}\approx\Hi^2/\Heq$ at the time of horizon
crossing, see eq. (3.13), massive modes give an order of magnitude of
fluctuations compatible with observations if \linebreak
$(\Hi/10^{16}$ {\sl GeV}$)^2\cdot(m/10^{-28}$ {\sl eV}$)^{1/2}=\Cal{O}(1)$.
For stable massive fields with $m\ge10^{-3}$ {\sl eV} this requires
$\Hi\le10^{10}$ {\sl GeV}, {\it i.e.} the energy density during inflation
has to fulfil \linebreak
$\ri\le(3\cdot10^{14}$ {\sl GeV}$)^4$.

\noindent{\bf4.2.} {\bfsl Unstable massive fields}\br\indent
In particle physics all massive bosons are unstable. They will decay into
lighter or massless particles. We consider the following simple scenario:
The massive scalar field decays with a rate $\Gamma$ into
ultra--relativistic particles, {\it i.e.} the scalar field has a mean
lifetime $\Gamma^{-1}$. In addition to the two horizon crossings and the
time when the mode becomes non--relativistic, there is now a fourth
important point in the history of the mode, namely the time of decay, where
$\Gamma\approx\H$. We have included this point in fig. 4.1. The decay does
not alter the spatial distribution of the energy density on large scales.
The decay products immediately decohere and thermalize, and they will behave
like classical radiation. Therefore after the decay the fluctuations
$\de\rho$ in the energy density of the massless, thermalized decay products
decrease in time as $a^{-4}$ in external FRW space--time.

Instead of evolving in time a true measure of the energy fluctuations in the
energy density of the decay products, like for instance the two--point
correlation function, we will give a quick estimate for their magnitude. We
continue the energy fluctuations $\dek$ from the time of decay by
redshifting with $a^{-1}$, as if the massive mode had transformed into
a massless (sub--horizon) mode at $\Gamma=\H$. This scenario is plotted in
{\bfsl fig. 4.3}. In this case we obtain the ratio
$$
{(\dek)_{m\neq0,\,\Gamma\neq0}\over(\dek)_{m=0}}\,\,=\,\,
                          \sqrt{{m\over\Gamma}}\quad(\hbox{for}\,\,\kph=\H).
\tag4.8
$$
The ratio is scale independent at the reference point $\kph=\H$, and does
not change afterwards. Therefore we can conclude that the energy
fluctuations $\de\rho$ in the massless, thermalized decay products have the
same scale dependence as in the case of a massless, fully coherent scalar
field. But their magnitude is enhanced over the massless case by a factor
$\sqrt{m\,\Gamma^{-1}}$ in the decay scenario. This simple estimate
indicates that with an unstable massive scalar field the scale invariant
(Harrison--Zel'dovich) fluctuations are preserved, and the small magnitude
$(\Hi/\Mp)^2$ of the massless case can easily be enhanced by a factor of ten
or more.

\vskip1truecm

\noindent{\ti 5. The Cosmic Variance}
\vskip0.5truecm
\noindent{\bf5.1.} {\bfsl The variance of $\,\,\ors$}\br\indent
In a finite sample $S$ with volume $V_S$ and mean linear extension
$\ell_S:=\root{\ssize3}\of{V_S}$, {\it i.e.} a finite patch of the universe,
the quantum fluctuations of the operator $E_S$ of the total energy,
$E_S:=\int_S d^3x\,\,\rho(x)$, play a crucial role. (In this section we
choose $a\equiv1$ at the time for which the analysis is performed.) For an
infinite sample $S$ the fluctuations of $E_S$ are negligible compared to the
expectation value $\langle E_S\rangle$, but in a finite $S$ the variance
$(\des/E_S)^2$ is relevant.

It is crucial to distinguish between spatial averages and ensemble averages
({\it i.e.} quantum mechanical expectation values). The spatial averages of
an operator $A$ will be denoted by a bar, $\overline A$, while for the
ensemble average brackets will be used, $\langle A\rangle$. We note that the
spatial average of the operator $\rho(x)$ over the sample $S$, {\it i.e.}
$\ors$, is proportional to the operator $E_S$,
$$
\ors\,:=\,\,\int_S {d^3x\over V_S}\,\,\rho(x)\,\,=\,\,{E_S\over V_S}.
\tag5.1
$$
The ensemble average $\rx$ of the operator $\rho(x)$ must be
$x$--independent in a cosmological context, therefore
$\r=\langle\ors\rangle$. The variance of the total energy in the sample is
$(\des)^2=\langle E_S^2\rangle-\langle E_S\rangle^2$, and the variance of
the average energy density in $V_S$ is
$$
(\dors)^2\,\,=\,\,\langle\ors^2\rangle-\langle\ors\rangle^2.
\tag5.2
$$
For a spherical sample of radius $R$ the variance $(\dors)^2$ is the same as
$\rt^2\ddR$ for a top hat window function, $W_R(\vx)\equiv1$ for
$|\vx|\le R$ and $W_R(\vx)\equiv0$ otherwise.

Consider now the (equal time) autocorrelation function $C(\ell)$, eq. (2.5).
The spatial (double) integral of $C(x,x')\equiv C(\ell)$ over the sample
yields the variance of $E_S$,
$$
\int_S d^3x\int_S d^3x'\,\,C(\ell)\,\,=\,\,
           \langle E_S^2\rangle-\langle E_S\rangle^2\,\,\equiv\,\,
                                                       V_S^2\cdot(\dors)^2.
\tag5.3
$$
In the limit $V_S\to\infty$ the boundary effects of $S$ become negligible,
and we obtain
$\int_S d^3x\int_S d^3x'\,C(\ell)\approx V_S\cdot\int_S d^3\ell\,C(\ell)$.
Thus we can rewrite eq. (5.3) as
$$
\left.
\aligned
(\des)^2\,\, &\simeq\,\,V_S\,\int_S d^3\ell\,\,C(\ell) \\
(\dors)^2\,\,&\simeq\,\,{1\over V_S}\,\int_S d^3\ell\,\,C(\ell)
\endaligned
\quad\right\}
\qquad\hbox{if}\quad V_S\to\infty.
\tag5.4
$$
This is equivalent to $(\dors)^2\simeq V_S^{-1}\,C(q=0)$ for $V_S\to\infty$.
Since $C(q=0)$ in our class of models is a positive constant, the absolute
fluctuations $\des$ increase with $\sqrt{V_S}$, while the relative
fluctuations $\dors$ decrease with $1/\sqrt{V_S}$. In the standard treatment
of the inflaton fluctuations only the contributions of the interference
terms between background and fluctuating part are kept, and $\dors$
decreases as $\ell_S^{-2}$ for large samples. Because of these different
asymptotic scaling laws the fluctuations in our model inevitably dominate
over the fluctuations in the inflaton scenario on large enough scales. Our
model predicts $\int d^3\ell\,C(\ell)=11\Hi^4\H/420\pi$. Dividing by the
total energy density $\rt$ we obtain
$$
{\dors\over\rt}\,\,\simeq\,\,\sqrt{{11\pi\over105}}\,{4\Hi^2\over3\Mp^2}\,
                       \bigg({\ell_{\H}\over\ell_S}\bigg)^{\tsize{3\over2}}
                             \quad\hbox{for}\,\,\ell_S\gg\ell_{\H}:=\H^{-1}.
\tag5.5
$$

\noindent{\bf5.2.} {\bfsl The finite sample correlation function}\br\indent
We define the operator
$$
\Drs(x)\,:=\,\,\rho(x)-\ors,
\tag5.6
$$
the difference between $\rho(x)$ and its spatial average in the sample $S$.
We consider the (equal time) autocorrelation function $C_S(\ell)$ of the
operator $\Drs$,
$$
C_S(\ell)\,:=\,\,\bla\Drs(\vx)\Drs(\vxp)\bra.
\tag5.7
$$
In a cosmological context $C_S(\ell)$ is only a function of the physical
separation $\ell$ of the two points. The (double) integral of $C_S(\ell)$
over the sample vanishes by definition,
$\int_S d^3x\int_S d^3x'\,C_S(\ell)\equiv0$, therefore the correlation
function $C_S(\ell)$ must change its sign at some value $\ell$ in the sample
(or at several values). The correlation function $C_S(\ell)$, which by
definition has at least one zero, is the experimentally accessible one, see
ref. [13]. But $C(\ell)$, which in our model has turned out to be strictly
positive, contains the ensemble average $\r$, which is unmeasurable (since
only one sample $S$ is available).

For small separations, $\ell\ll\ell_S$, the two correlation functions are
related as
$$
C(\ell)\,\,\simeq\,\,C_S(\ell)+(\dors)^2\,\,=\,\,
                           C_S(\ell)+{1\over V_S}\,\int d^3\ell\,\,C(\ell)
                                           \quad\hbox{for}\,\,\ell\ll\ell_S,
\tag5.9
$$
their difference is $\ell$--independent and equal to the cosmic variance.
For any fixed $\ell$ the two correlation functions are identical in an
infinite sample. In our model the zero of $C_S(\ell)$ moves to infinity as
$V_S\to\infty$.

\vskip1truecm
We would like to thank Christophe Massacand and Slava Mukhanov for
many valuable discussions and comments.

\vfill\eject

\noindent
{\ti References}
\vskip0.5truecm
\item{[1]}  N.D. Birrell and P.C.W. Davies, {\sl Quantum Fields in Curved
             Space}, Cambridge, 1982.
\item{[2]}  K. Olive, {\sl Phys. Rep.} {\bf 190}(1990), 307; \br
            V.F. Mukhanov, H.A. Feldman, and R.H. Brandenberger,
             {\sl Phys. Rep.} {\bf 215}(1992), 203.
\item{[3]}  E. Kolb and M.S. Turner, {\sl The Early Universe},
             Addison--Wesley, 1990;\br
            A.D. Linde, {\sl Particle Physics and Inflationary Cosmology},
             Harwood Academic, 1990.
\item{[4]}  G.F. Smoot, {\it et al.}, {\sl Ap. J.} {\bf 396}(1992), L1; \br
            E.L. Wright, {\it et al.}, {\sl Ap. J.} {\bf 396}(1992), L13.
\item{[5]}  A.H. Guth and S.-Y. Pi, {\sl Phys. Rev. Lett.} {\bf 49}(1982),
             1110; \br
            S.W. Hawking, {\sl Phys. Lett.} {\bf 115B}(1982), 295; \br
            A.A. Starobinsky, {\sl Phys. Lett.} {\bf 117B}(1982), 175; \br
            J.M. Bardeen, P.J. Steinhardt and M.S. Turner,
             {\sl Phys. Rev. D} {\bf 28}(1983), 679.
\item{[6]}  H.F. M\"uller and C. Schmid, {\sl ETH--Preprint} ETH--TH/93--15,
             gr--qc/9401020.
\item{[7]}  R. Durrer and M. Sakellariadou, {\sl Z\"urich University
             Preprint} ZU--TH/9--94, \linebreak astro--ph/9494043.
\item{[8]}  H.F. M\"uller and C. Schmid, {\sl ETH--Preprint} ETH--TH/94--35.
\item{[9]}  D. Seckel and M.S. Turner, {\sl Phys. Rev. D} {\bf32}(1985),
            3178.
\item{[10]} L.M. Krauss and M. White, {\sl Phys. Rev. Lett.} {\bf69}(1992),
            869.
\item{[11]} B. Allen, {\sl Phys. Rev. D} {\bf 37}(1988), 2079.
\item{[12]} H.F. M\"uller, ETH--dissertation No. 10727 (1994), 125 pp.
\item{[13]} P.J.E. Peebles, {\sl The Large--Scale Structure of the
             Universe}, Princeton, 1980.

\vfill\eject

\noindent
{\ti Figure Captions}
\vskip1truecm
\noindent{\bfsl Fig. 3.1:}
  Time dependence of the energy fluctuations $\dek$ for $m=0$
\vskip0.5truecm

\noindent{\bfsl Fig. 3.2:}
  The energy--energy correlation function $C(\ell)$ at equality time. Two
  domains: On scales $\ell\ll\laeq=\Heq^{-1}=13(\Om h^2)^{-1}$ {\sl Mpc} a
  logarithmic increase towards smaller $\ell$. For $\ell\gg\laeq$ a
  fall--off as $\ell^{-4}$. The cusp in the curve is at
  $\ell_{\rm cusp}=2\laeq$.
\vskip0.5truecm

\noindent{\bfsl Fig. 3.3:}
  The variance $(\de\rho_R)^2:=\rt^2\ddR$ as a function of the smearing
  width $R$ at equality time. Two domains: For
  $R\ll\laeq=\Heq^{-1}=13(\Om h^2)^{-1}$ {\sl Mpc} a logarithmic increase
  towards smaller $R$. On scales $R\gg\laeq$ a fall--off as $R^{-3}$.
\vskip0.5truecm

\noindent{\bfsl Fig. 3.4:}
  The energy--energy correlation function $C(q)$ at equality time. Two
  domains: On scales $q\gg\Heq=(13(\Om h^2)^{-1}$ {\sl Mpc}$)^{-1}$ a
  fall--off as $q^{-3}$. For super--horizon wavelengths, $q\ll\Heq$, $C(q)$
  is constant.
\vskip0.5truecm

\noindent{\bfsl Fig. 4.1:}
  Time development of the physical length scales. Three relevant scales for
  a stable massive mode: the Hubble parameter $\H$, the mass $m$, and the
  energy $\eph$. A fourth scale for an unstable field, the decay rate
  $\Gamma$. Three important points: the first horizon crossing when
  $\eph\simeq\kph=\Hi$, the time when the mode becomes non--relativistic,
  $\kph=m$, and the second horizon crossing when $\eph\simeq m=\H$. For an
  unstable particle a fourth important point where $\Gamma=\H$. The point
  $\kph=\H$ is dynamically irrelevant for massive modes.
\vskip0.5truecm

\noindent{\bfsl Fig. 4.2:}
  Time evolution of the energy fluctuations $\dek$ for a stable massive
  field with $0<m\ll\Hi$.
\vskip0.5truecm

\noindent{\bfsl Fig. 4.3:}
  Time evolution of the energy fluctuations $\dek$ for an unstable massive
  field with $0<m\ll\Hi$ and lifetime $\Gamma^{-1}$. Comparison between the
  fluctuations $\dek$ for unstable massive (solid) and massless scalars
  (dashed). At $\kph=\H$ we see the enhancement.

\end